# Community-developed checklists for publishing images and image analyses

Christopher Schmied[1,2], Michael S. Nelson[3], Sergiy Avilov[4], Gert-Jan Bakker[5], Cristina Bertocchi[6,7], Johanna Bischof[8], Ulrike Boehm[9], Jan Brocher[10], Mariana Carvalho[11], Catalin Chiritescu[12], Jana Christopher[13], Beth A. Cimini[14], Eduardo Conde-Sousa[15], Michael Ebner[1], Rupert Ecker[16, 17,18], Kevin Eliceiri[19], Julia Fernandez-Rodriguez[20], Nathalie Gaudreault[21], Laurent Gelman[22], David Grunwald[23], Tingting Gu[24], Nadia Halidi[25], Mathias Hammer[23], Matthew Hartley[26], Marie Held[27], Florian Jug[1], Varun Kapoor[28], Ayse Aslihan Koksoy[29], Judith Lacoste[30], Sylvia Le Dévédec[31], Sylvie Le Guyader[32], Penghuan Liu[33], Gabriel G. Martins[34], Aastha Mathur[8], Kota Miura[35], Paula Montero Llopis[36], Roland Nitschke[37], Alison North[38], Adam C. Parslow[39], Alex Payne-Dwyer[40], Laure Plantard[22], Rizwan Ali[41], Britta Schroth-Diez[42], Lucas Schütz[43], Ryan T. Scott[44], Arne Seitz[45], Olaf Selchow[46], Ved P. Sharma[38], Martin Spitaler[47], Sathya Srinivasan[48], Caterina Strambio-De-Castillia[49], Douglas Taatjes[50], Christian Tischer[51] and Helena Klara Jambor[52]

*Affiliations: see end of manuscript.*

## Abstract

Images document scientific discoveries and are prevalent in modern biomedical research. Microscopy imaging in particular is currently undergoing rapid technological advancements. However for scientists wishing to publish the obtained images and image analyses results, there are to date no unified guidelines. Consequently, microscopy images and image data in publications may be unclear or difficult to interpret. Here we present community-developed checklists for preparing light microscopy images and image analysis for publications. These checklists offer authors, readers, and publishers key recommendations for image formatting and annotation, color selection, data availability, and for reporting image analysis workflows. The goal of our guidelines is to increase the clarity and reproducibility of image figures and thereby heighten the quality and explanatory power of microscopy data is in publications.

## Introduction

Images and their analyses are widespread in life science and medicine. Microscopy imaging is a dynamic area of technology development, both in terms of hardware and software. This is especially true in the area of light microscopy with great recent improvements in sensitivity, and spatial and temporal collection. Resources developed by scientists help researchers to navigate designing microscopy experiments and obtaining image data (Brown, 2007; North, 2006; Senft et al., 2022), and cover aspects such as sample preparation (North, 2006), microscope usage (Jonkman, 2020; North, 2006), method reporting (Hammer et al., 2021; Heddleston et al., 2021; Montero Llopis et al., 2021; Rigano et al., 2021), or fluorophore and filter usage (Kiepas et al., 2020; Laissue et al., 2017). Despite widespread adoption of microscopy as a tool for biology and biomedical research, the resulting image figures in publications at times fail to fully communicate results or are not entirely understandable to audiences. This may be because authors do not include comprehensive imaging method statements Sheen et al., 2019), or because they omit basic information in figures such as specimen size or color legends (Jambor et al., 2021), which are key to fully understanding the data. To ensure that images are presented in a clear, standardized, and reproducible manner, it is essential that the scientific community establishes unified and harmonized guidelines for image communication in publications.

Images document biological samples and ranges of their phenotypes. Increasingly, microscopy images are also a source of quantitative biological data where and variables are measured with a growing number of image analysis software packages (FIJI/ImageJ Schindelin et al., 2012), CellProfiler (Stirling et al., 2021), KNIME (Dietz et al., 2020), commercial software packages such as ZEN Blue, Amira, Imaris, Arivis, and Python software libraries (Perkel, 2021), https://scikit-image.org/; see also Eliceiri 2012). Image analysis is often a workflow of many steps, such as image reconstruction, pre-processing, segmentation, post-processing, rendering, visualization and statistical analysis, many of which require expert knowledge (Aaron and Chew, 2021; Miura and Tosi, 2017). A comprehensive publication of quantitative image data then not only includes basic specimen and imaging information but additionally the image processing and analysis steps that produced the data plot and statistics. Towards fully reproducible image analysis it is also essential that images and workflows are available to the community, e.g., in image repositories or archives (Ellenberg et al., 2018; Hartley et al., 2022; Williams et al., 2017), and code repositories such as Github (Ouyang et al., 2022).

To ensure that image figures provide insights to their readership, any supportive experimental metadata and image analysis workflows must be clear and understandable ("what is the pixel size", "what does the arrow mean"), accessible ("are colors visible to colorblind audiences"), representative (no cherry picking), and reproducible ("how were the data processed", "can one access and re-analyze the images"). In the framework of the initiative for 'Quality Assessment and Reproducibility for Instruments and Images in Light Microscopy', QUAREP-LiMi (Boehm et al., 2021; Nelson et al., 2021), the 'Image Analysis and Visualization workgroup' established community consensus checklists to help scientists publish understandable and reproducible light microscopy images and image analysis procedures. Where applicable, the checklists are aligned with the FAIR principles, which were developed as recommendations for research data (Findability, Accessibility, Interoperability, and Reusability) (Wilkinson et al., 2016).

## Scope of checklists

The scope of the checklists is to help scientists publish fully understandable and interpretable images and results from image analysis (Figure 1). In this work the term images includes raw or essentially unprocessed light microscope data, compressed or reconstructed images, and quantification results obtained through image analysis (See Glossary). While the focus of QUAREP-LiMi is on light microscopy images in life sciences, the principles may also apply to figures with other images (photos, electron micrographs, medical images) and to image data beyond life sciences. The intended audience of the checklists are novices or non-experts occasionally using light microscopy, and also experts (core facility staff, global bioimage community) who review image data or teach image handling.

The checklists do not include principles for designing imaging experiments and recommendations to avoid image manipulation. Previous literature covers experimental design for microscopy images, including truthful image acquisition and ensuring image quality (Brown, 2007; Faklaris et al., 2022), examples and recommendations for avoiding misleading images (Bik et al., 2018, 2016; Cromey, 2013; CSE, 2012; North, 2006; Rossner and Yamada, 2004), detection of image manipulation (Bucci, 2018; Koppers et al., 2017; Van Noorden, 2022, 2020), appropriate image handling and analysis (Aaron and Chew, 2021; Hammer et al., 2021; Martin and Blatt, 2013; Miura and Norrelykke, 2021), guidelines for writing materials and methods sections for images (Marques G, 2020), and recommendations for general figure preparation Nature Guidelines). These topics are therefore not covered in the checklists.

The checklists cover image (Figure 2, Suppl. Figure 1) and image analysis (Fig. 8, Suppl. Figure 2) and are structured into three levels that prioritizes legibility and reproducibility.

- The first reporting level ("Minimal") describes necessary, non-negotiable requirements for the publication of image data (microscopy images, data obtained through image analysis). Scientists can use these minimal criteria to identify crucial gaps before publication.
- The second reporting level ("Recommended") defines measures to ensure the understandability of images and aims to reduce the efforts toward evaluating image analysis . We encourage scientists to aim for the "Recommended" level as their image publication goal. However, we acknowledge that some aspects (e.g., large data in repositories) may today be still unattainable for some authors.
- The third reporting level ("Ideal") are recommendations we encourage scientists to consider adopting in the future.

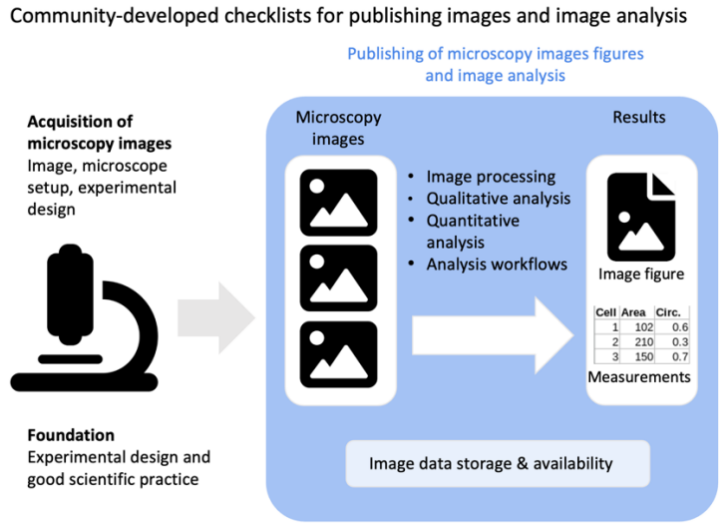

*Figure 1. Scope of the checklists of image figure and image analysis requirements. The checklists present easy-to-use guidelines for publishing microscopy image figures and image analysis workflows.*

## Checklists for image publication

**IMAGE FORMATTING.** Preparing a figure begins with the selection of representative images from the dataset. When quantitative measurements are reported in a chart, an example of the input image should be shown; when ranges of phenotypes are described, several images may be necessary for illustrating the diversity. To quickly focus the audience on key structures in the image, it is permitted to crop areas without data or with non-relevant data (Figure 3A). As a rule, cropping, similar to selecting the field-of-view on the microscope, is allowed as long as this does not change the meaning of the conveyed image. Image rotation may help standardize specimen orientation (e.g., apical side of cells upwards) and is permitted. Image rotation by angles different from 90 degrees or multiples thereof, however, changes the intensity values through interpolation and therefore alters the information in the image (Cromey, 2013, 2010; Schmied and Jambor, 2020). When cropping and rotating, authors should ensure that the operation does not affect the original information contained in the image, and quantifications, especially intensity measurements, should be performed beforehand (Miura and Norrelykke, 2021). Overall, any loss in image quality may be acceptable for image figure preparation, but image quantifications should be done beforehand. In a figure, individual images should be well separated (spacing, border, see Figure 3B) to avoid misleading image-splicing (Bik et al., 2016; Cromey, 2013).

When presenting two magnifications of the same image (e.g., a full- and a zoomed/inset view), the position of the inset in the full-view image should be made clear; if the inset is placed on top of the full-view image, e.g., to save space, it should not obstruct key data (Figure 3C). If an inset is digitally zoomed, the original pixels should not be interpolated but "resized" to maintain the original resolution. Overall, the image should be sufficient in size so that audiences can identify all relevant details. Limitations for figure height/width may be set by publishers with figures commonly required at 300 dots per inch (dpi) or pixels per inch (ppi) resolution. Please refer to existing journal guidelines for further information.

## Checklist for image publishing

### Image format

| | | Minimal | Recommended | Ideal |
|---|---|---|---|---|
| 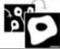 | Focus on relevant image content (e.g. crop, rotate, resize) | ☐ | ☐ | ☐ |
| 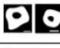 | Separate individual images | ☐ | ☐ | ☐ |
| 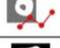 | Show example image used for quantifications | ☐ | ☐ | ☐ |
| 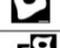 | Indicate position of zoom-view/inset in full-view/ original image | ☐ | ☐ | ☐ |
| 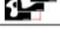 | Show images of the range of described phenotype | ☐ | ☐ | ☐ |

### Image colors and channels

| | | Minimal | Recommended | Ideal |
|---|---|---|---|---|
| 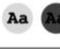 | Annotation of channels (staining, marker etc.) visible | ☐ | ☐ | ☐ |
| 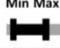 | Adjust brightness/contrast, report adjustments, use uniform color-scales | ☐ | ☐ | ☐ |
| 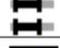 | Image comparison: use same adjustments | ☐ | ☐ | ☐ |
| 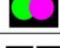 | Multi-color images: accessible to color blind | ☐ | ☐ | ☐ |
| 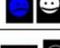 | Channel color high visibility on background | ☐ | ☐ | ☐ |
| 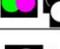 | Provide grey-scale for each color channel | | ☐ | ☐ |
| 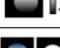 | Provide color scales for intensity values (greyscale, color, pseudo color…) | | ☐ | ☐ |
| 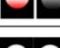 | Pseudo-colored images: additionally provide greyscale version for comparison. | | | ☐ |
| 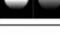 | Gamma adjustments: additionally provide linear-adjusted image for comparison | | | ☐ |

### Image annotation

| | | Minimal | Recommended | Ideal |
|---|---|---|---|---|
| 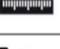 | Add scale information (scale bar, image length) | ☐ | ☐ | ☐ |
| 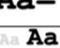 | Explain all annotations | ☐ | ☐ | ☐ |
| 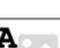 | Legible annotations (point size, color) | ☐ | ☐ | ☐ |
| 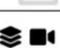 | Annotations should not obscure key data | ☐ | ☐ | ☐ |
| 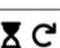 | Annotate image dimensions (z-distance in image stacks), image pixel size, imaging intervals (time-lapse in movies) | | ☐ | ☐ |
| 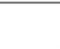 | Annotate imaging details e.g., exposure time | | | ☐ |

### Image availability

| | | Minimal | Recommended | Ideal |
|---|---|---|---|---|
| 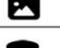 | Images are shared (lossless compression/microscope images) | ☐ | ☐ | ☐ |
| 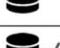 | Image files are freely downloadable (public database) | | ☐ | ☐ |
| 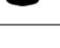 | Image files are in dedicated image database (added value database or image archive) | | | ☐ |

*Figure 2. Checklist for image publication including points to be addressed on image format, image colors and channels, image annotations, and image availability.*

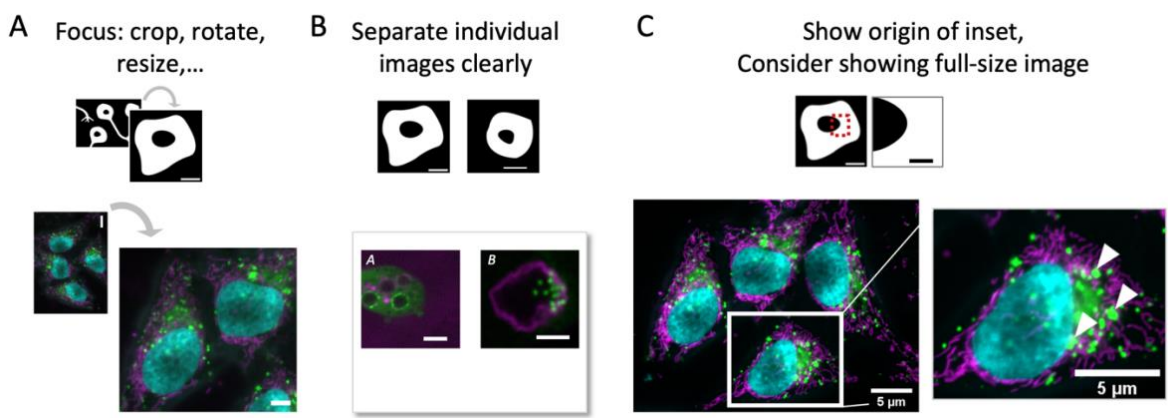

*Figure 3. Image formatting may include (A) image cropping, rotation, and resizing, (B) image spacing in the figure, and (C) presenting several magnifications (zoom, inset) of images.*

**IMAGE COLORS AND CHANNELS.** Fluorescent light microscopes use a range of wavelengths to generate images of specimens. In the images, the light intensity for individual wavelengths, most often in grayscale, is assigned or mapped to a visible color scheme. In multi-colored images, several channels are overlaid to compare data from several channels.

Microscopy images often must be processed to adapt the bit depth to the visible range (Brown, 2007; Russ, 2006). Usually, brightness/contrast is adjusted for each channel independently in many software platforms (e.g., ImageJ/FIJI) by defining the minimum and maximum displayed intensity values before converting these into 8-bit (for screen display, printing). Intensity range adjustments should be monitored with the e.g. the image histogram and done with care: a too wide intensity range results in 'faded' images that lack details, while a too narrow intensity range removes data (Figure 4A). Scientists must be especially attentive with auto-contrast/auto-level, image intensity normalization, non-linear adjustments ('gamma,' histogram equalization, local contrast e.g., CLAHE, Zuiderveld, 1994), image filters, and image restoration methods e.g., deconvolution, Noise2Void, CARE, etc. (Fish et al., 1995; Krull et al., 2019; Richardson, 1972; Weigert et al., 2018) as their improper application may result in misleading images. When images are quantitatively compared in an experiment, the same adjustments and processing steps must be applied. If deemed critical for understanding the image data, advanced image processing steps (e.g., deconvolution, Noise2Void, CARE) may need to be indicated in the figure (figure legend), in addition to the material and methods sections.

Next, image colors must be interpretable and accessible to readers, and not mislead (Crameri et al., 2020). For full-color (e.g., histology) images, the staining/preparation method, and for fluorescence microscope images the channel-specific information (fluorophore/labeled biomolecule) should be annotated (Figure 4B, also see next section). In fluorescence microscope images, the channels can be assigned a user-defined color scheme, often referred to as lookup table (LUT), which should be chosen such that imaged structures are well distinguishable from the background and accessible to color-blind audiences (Jambor et al., 2021). Grayscale color schemes will allow the audience to interpret image details best since they are uniformly perceived, which allows unbiased interpretation. Inverting image LUTs, to display intensities on a white instead of a black background may enhance signal contrast further, but be aware that different software handles this calculation differently.

A few steps may overall improve the understandability of colors. For multi-colored fluorescent images, consider if showing individual channels in separate, grayscale images may help readers fully appreciate details (Figure 4C). A separate, linear-adjusted grayscale version may help when images were adjusted with non-linear adjustments or pseudo-colored LUTs (e.g., 'jet,' 'viridis,' and 'union-jack'), which map intensity values to a dual- or multiple color sequential scheme. Annotation of intensity values with a color scale bar ('calibration bar') helps to orient readers and is essential for pseudo-colored and non-linear color schemes (Figure 4D). Calibration bars should indicate absolute intensity values to informs audiences about the displayed intensity range and can be prepared with Imaris and ImageJ/FIJI (see ImageJ user guide).

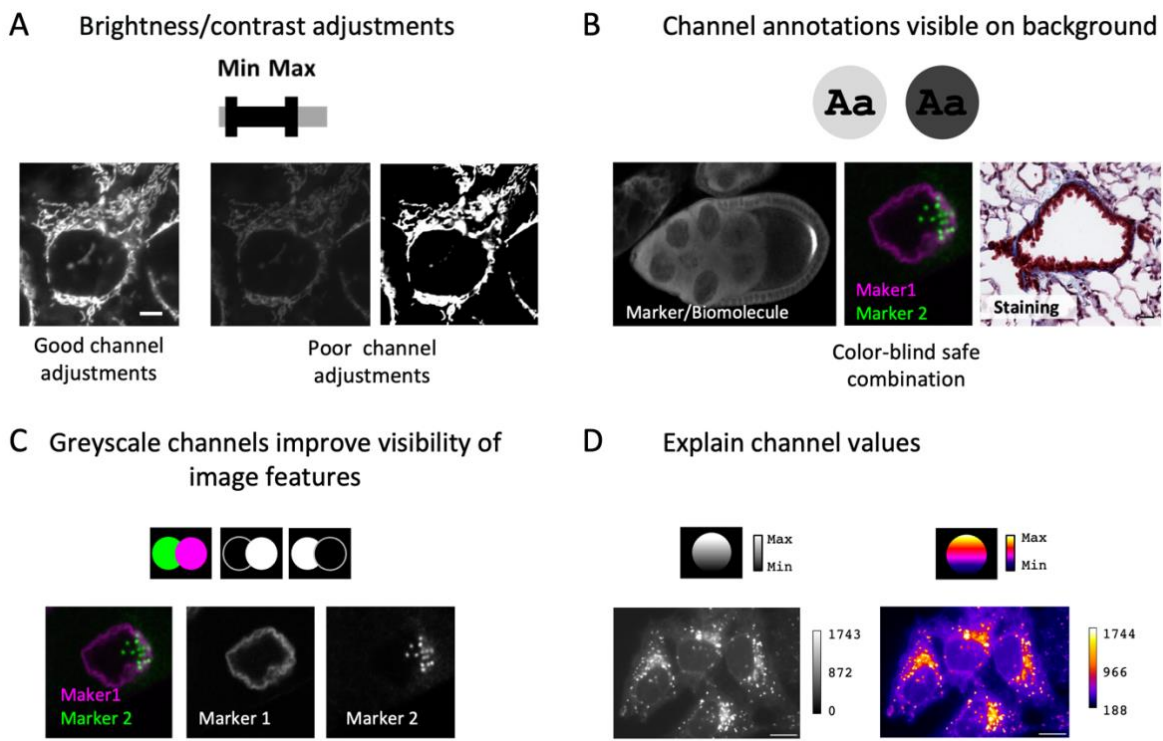

*Figure 4. Image colors and channels. (A) Adjust brightness/contrast to achieve good visibility of the imaging signal. (B) Channel information should be annotated and visible to audiences (high contrast to background color, visible to color-blind audiences). (C) Image details are most unbiased in grayscale. (D) It is best practice to publish legends to color scales with images, and mandatory for pseudo-color scales.*

**IMAGE ANNOTATION.** Light microscopy images show objects sized from submicron to millimeters resolution. As physical size is not obvious without context, annotating the scale for publication is therefore necessary. Including a scale bar of a given size (in or next to image) is needed to orient audiences (Figure 5A). The corresponding size statement/dimension, e.g., "0.5 mm", can be placed next to the scale bar (when not possible then in the figure legend). To avoid quality changes (pixelated/illegible text) when adapting (re-size, compress) figures for publication, annotations should be added as vector graphics. Statements about the physical length of the entire image are acceptable alternatives to scale bars. Magnification statements should be avoided as pixel size can be determined by a number of factors e.g., sampling rate or binning, and does not only depend on the objective magnification.

Many images include further annotations such as symbols (arrows, asterisks), letter codes, or regions-of-interest (dashed shapes) which must be explained in the figure or figure legend (Figure 5B). Annotations placed on top of images should not obscure key image data and must be legible (font, font size). Furthermore, annotations should have good visibility on the image, i.e., must be legible to color-blind persons and distinguishable from image background and image content (annotation shapes/colors distinct from object shapes/colors). Symbols that resemble image data should be avoided, and note that symbols with clear vertical/horizontal arrangement are easier to distinguish than randomly oriented symbols on busy backgrounds (for examples see: (Jambor et al., 2021). Images without annotations should be available to audiences (see Image Availability). When images are used for quantitative measurements (length, volume, time constants) it is advisable to include statements about the sampling rate in space/time. Depending on the situation, this may be the pixel-size for 2D imaging, z-distance for image volumes/3D-images, or imaging frequency for time-lapse data (Figure 5C).

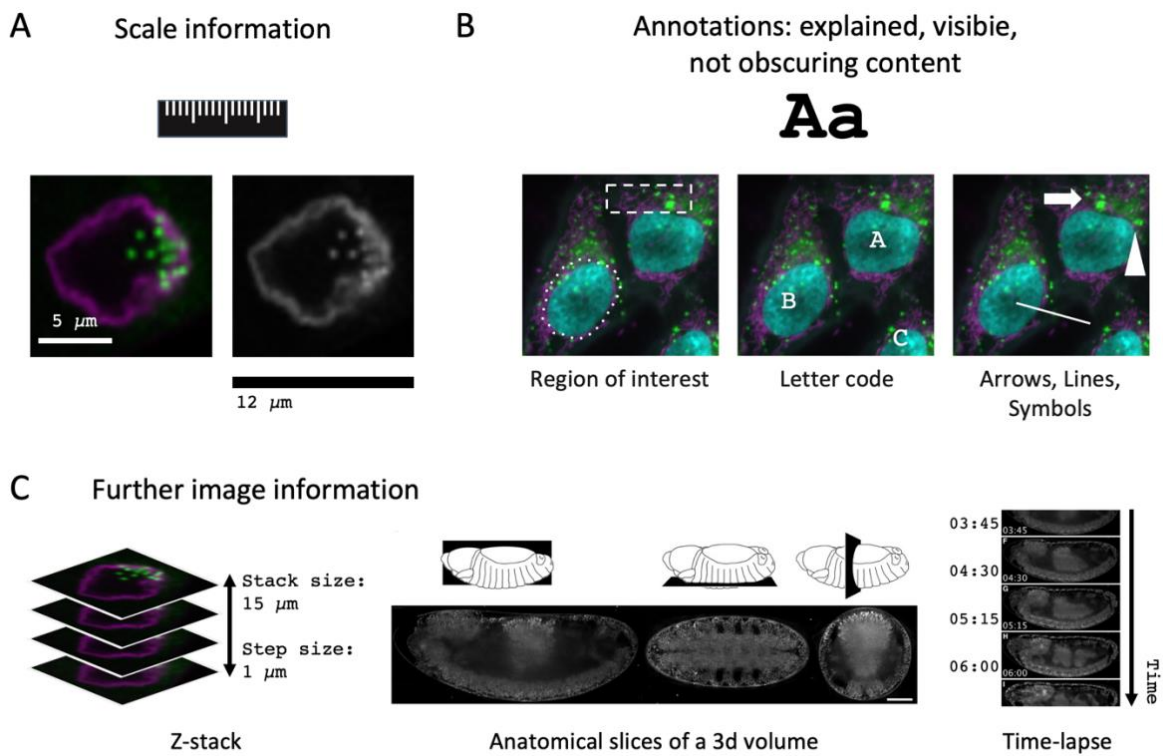

*Figure 5. Image Annotation. (A) Possible ways to provide scale information. (B) Features in images can be annotated with symbols, letters, or region-of- interest. (C) For advanced image publication, information on z-stack or voxel size, anatomical view, or camera settings such as pixel-dwell time, exposure time, or frame-time may be required.*

**IMAGE AVAILABILITY.** Any image processing should be performed on a duplicate copy of the original microscope image (Cromey, 2010; Schmied and Jambor, 2020) and upon publication both, the original image (or a lossless compressed version) and the published image should be available. The specific file type of the original image depends on the microscope type and the vendor. The definition of 'original data' or 'raw data', and whether its storage is feasible, depends on the specific microscopy technique. In data-heavy techniques that collect sparse information (e.g., light-sheet microscopy, time-lapse), reconstructed images may faithfully capture the key data and should be made available. To retain the metadata, a conversion into open formats such as OME-TIFF (Linkert et al., 2010; which supports uncompressed, lossless, but also lossy compressed files) is compatible with broad applications to allow re-analysis of image data. If only a compressed version may be kept (i.e., a file in which image channels and annotations are irretrievably merged), PNG files are superior to the JPEG format as they allow lossless compression (Cromey, 2010).

As a minimal requirement, image files shown in figures or used for quantification should be available. When possible (see limitations above), lossless compressed files which allow replication of the analysis workflow should be shared or made available (Figure 6). We strongly discourage that authors make images available "upon request" since this has been shown to be inefficient (Gabelica et al., 2022; Tedersoo et al., 2021), however at present infrastructure is not sufficiently in place to ban this option. A clear advancement is depositing both the published and the original images in a public repository with an open license that permits re-use in the scientific context (CC-BY, CC0). Zenodo, OSF, figshare are current options also for image data, however these have file size limitations. OMERO servers (https://www.openmicroscopy.org/omero/institution/) enable institutions but also individual labs to host public or private (access controlled) image sharing databases (Overview of current repositories, see Suppl. Figure 3). Long-term ("Ideal"), uploading of images with all experimental metadata to dedicated, specialized or fully searchable image databases has the potential to unlock the full power of image data for automated image and/or metadata searches, and the possibility of image data re-use. The databases allowing such functionalities and more include the BioImage Archive (a primary repository which accepts any image data in publications), the Image Data Resource (which publishes specific reference image datasets), or EMPIAR (a dedicated resource for Electron Microscopy datasets).

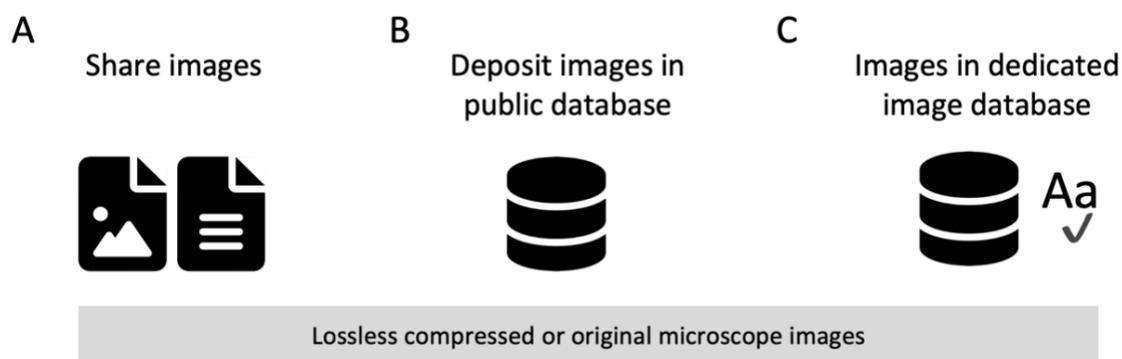

*Figure 6. Image Availability. (A) Currently, image data is often shared 'upon request'. (B) More images along with the image metadata should be available for download in public databases, and in the future (C) also archived in dedicated, added-value databases, in which images are machine searchable or curated.*

## Checklists for publication of image analysis workflows

Image analysis workflows usually combine several processing steps carried out in a specific sequence to mathematically transform the input image data into a result (i.e., image for visualization or data for a plot; Figure 7, Miura et al., 2020). As images are numerical data, image processing invariably changes these data and thus needs to be transparently documented (Cromey, 2013, 2010; Miura and Norrelykke, 2021). We developed separate checklists for scientists wishing to publish results originating from image processing and image analysis workflows (Figure 8, Suppl. Figure 2). Focusing on easy implementation of the checklists we propose three categories:

1. Established workflows or workflow templates: workflows available in the scientific literature or well established in the respective fields.

2. Novel workflows: established or new image analysis components (available in software platforms or libraries) are assembled by researchers into a novel workflow.
3. Machine learning (ML) workflows: ML uses an extended technical terminology and ML workflows that utilize deep neural networks ('deep learning') face unique challenges with respect to reproducibility. Given the rapid advancements in this field, we created a separate ML checklist.

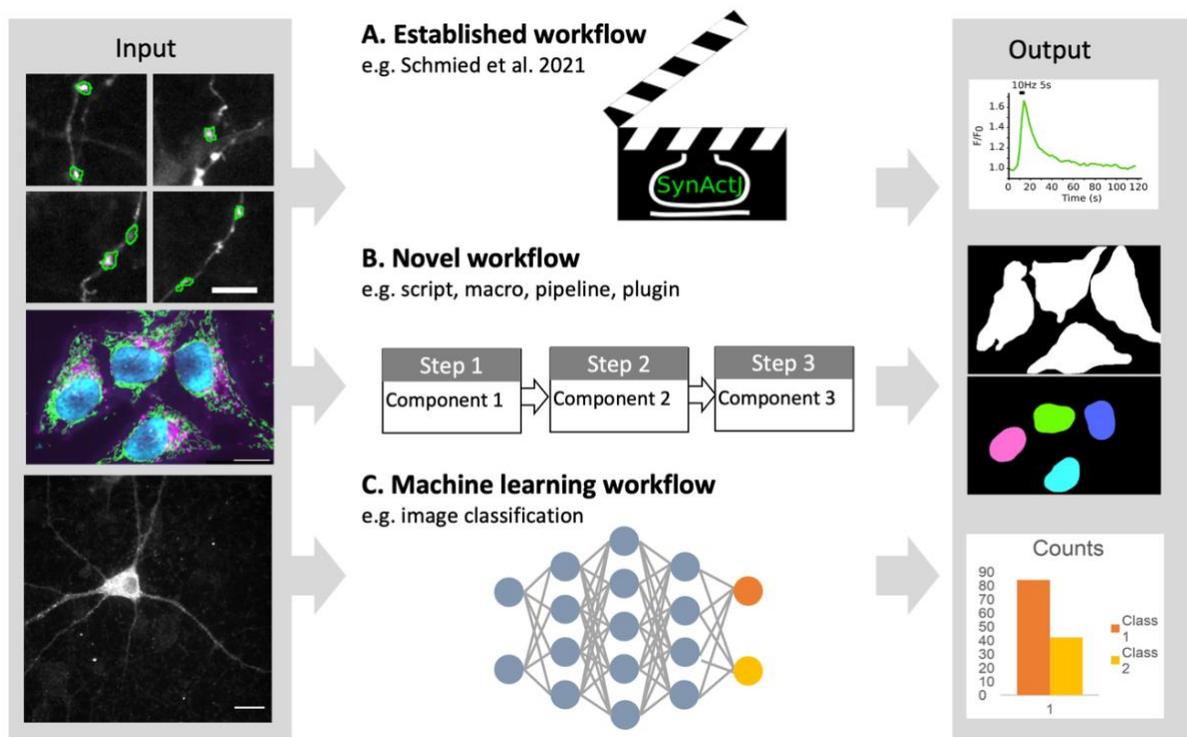

*Figure 7. Image analysis. (A) An established workflow template is applied on new image data to produce a result (plot). (B) A new sequence of existing image analysis components is assembled into a novel workflow for a specific analysis (image segmentation). (C) Machine learning workflows learn specific tasks from data, and the resulting model is applied to obtain results.*

**ESTABLISHED WORKFLOWS.** Examples of well-established workflows are published pipelines for cell profiler ([CellProfiler published pipelines](), [CellProfiler examples]()), workflows in KNIME (Fisch et al., 2018), specialized plugins and easy-to-use scripts in ImageJ (Erguvan et al., 2019; Klickstein et al., 2020; Schmied et al., 2021), tools and plugins that solve generic image analysis problems such as tracking (Tinevez et al., 2017) or pixel classification (Arganda-Carreras et al., 2017; Arzt et al., 2022). For these workflows extensive expertise, documentation, and tutorials already exist that allow others (e.g., reviewers, readers) to reproduce the workflow and to judge the validity of the results. Scientists publishing images or image analysis results processed with established workflows thus can focus on documenting key parameters only.

**Minimal.** The authors must cite the used workflow. The specific software platform or library needs to be cited if the workflow is not available as a stand-alone tool. Key processing parameters must be reported. To validate the performance of the workflow and its settings, example input and output data must be provided. Any manual interventions (e.g., ROIs) must be clarified.

**Recommended.** To ensure proper reproduction, the precise version numbers of the workflow and the platform used are vital and should be documented in the methods. If the used software does not allow the researcher to easily define and retrieve a specific versions number, the exact version used should be deposited as a usable executable or code. Authors should state all settings in the methods or the supplements of the article. Providing data upon request is an ineffective method for data sharing (Gabelica et al., 2022). Thus, authors should provide the example input, output and any manual regions of interest via a public repository (see above).

**Ideal.** Documenting the usage of software in the form of a screen recording or in the case of command line tools via reporting all executed commands in detail, greatly facilitates understanding of the workflow application and therefore reproduction. To avoid any variation arising from factors such as computer hardware or operating system authors could provide cloud-hosted solutions (Berginski and Gomez, 2013; Hollandi et al., 2020; Stringer et al., 2021)(kiosk-imagej-plugin) or the workflow packaged in a software container (docker, Singularity)(da Veiga Leprevost et al., 2017).

NOVEL WORKFLOWS. Novel image analysis workflows assemble components into a new sequence e.g., a macro in Fiji, a pipeline in CellProfiler or workflow in KNIME in an original way. To ensure reproducibility of the analysis, it is essential to report the specific composition and sequence of such novel workflows.

**Minimal.** The individual components utilized in the novel workflow must be cited, named and/or described in detail in the methods section along with the software platform used. It is essential tothat scientists specify or provide the exact software versions of the used components and software platform in the methods if possible. Authors must describe the sequence in which these components have been applied. Key settings (e.g., settings that deviate from default settings) must be documented in the methods section. Finally, the developed workflow must be shared as code (e.g., via code repositories https://github.com/), pipelines (e.g., KNIME workflow, CellProfiler pipeline) if possible, along with example input, output, and any manually generated inputs (i.e., ROIs), must be made available (See Image Availability). Novel workflows that were created using software that does not allow scripting, the workflow steps should be carefully described as a text.

**Recommended.** Disclose and describe all settings of the workflow to help the reproduction of the analysis. Provided example input, output, and manual inputs (ROIs) via public repositories such as Zenodo (European Organization For Nuclear Research and OpenAIRE 2013) . The developer should describe the rationale as well as the limitations of the workflow and the used components in more detail in the methods or supplements. Evidence of the adequacy and efficiency of the used algorithms on the published data and potentially even comparisons to related established workflows, when possible, facilitate such a documentation.

**Ideal.** To further promote reproducibility, add documentation such as a screen recording or a text-based tutorial of the application of the workflow. To enable the efficient reproduction of an analysis with a novel workflow, provide easy installs (e.g., update sites, packages) or easy software reproduction (e.g., via software containers), and easy-to-use user interfaces of software (i.e., graphical user interfaces). Publish the novel workflow as independent methods papers with extensive documentation and online resources (Arganda-Carreras et al., 2017; Arzt et al., 2022; Erguvan et al., 2019; Fisch et al., 2018; Klickstein et al., 2020; Schmied et al., 2021; Tinevez et al., 2017). Taken together, with extensive documentation, ease of installation and use will ultimately contribute to the novel workflow becoming well-established and reproduced within the community (a future established and published workflow template) (Cimini et al., 2020).

**MACHINE LEARNING WORKFLOWS.** Machine learning, and especially deep learning, have recently become capable of surpassing the quality of results of even the most sophisticated conventional algorithms and workflows and are continuing to advance (Laine et al., 2021). Deep learning procedures are quickly adapted to microscopy image tasks such as U-net (Ronneberger et al., 2015) for cell segmentation (Falk et al., 2019), Noise2Void for image reconstruction (Krull et al., 2019), StarDist (Schmidt et al., 2018; Weigert et al., 2020), (Schmidt et al., 2018; Weigert et al., 2018) Cellpose (Stringer et al., 2021) for instance segmentation, DeepProfiler (Moshkov et al., 2022) for feature extraction, and Piximi (https://www.piximi.app/) for image classification.

In machine learning workflows (supervised, unsupervised, self-supervised, shallow or deep learning), the input image data is transformed by one or multiple distinct mathematical operations into a scientific result. The instructions for this transformation are learned from provided data (e.g., labeled data for supervised learning, and unlabeled data for unsupervised learning) to produce a machine learning model. However, the precise makeup of this model is not easily accessible to a user and depends strongly on the quality and nature of the supplied training data as well as the specific training parameters. Biases in the training data/errors in the labels of ground truth for supervised machine learning will bias machine learning models (Larrazabal et al., 2020; Obermeyer et al., 2019; Seyyed-Kalantari et al., 2020). Reporting is thus even more critical for reproducibility and understandability when ML applications are applied for image analysis.

Three major approaches are widely used in ML-based image analysis today, which require different documentation : 1) pre-trained models are directly applied to new image data and referral to existing references is sufficient (minimal). 2) pre-trained models are re-trained (transfer learning) with novel image data to improve the application, and in this case more information must be provided (Recommended). 3) models are trained de-novo, in which case extensive documentation is required for reproducibility (Ideal).

**Minimal.** The precise machine learning method needs to be identifiable. Thus, the original method must be cited. At the minimum, access to the model that has been produced in the particular learning approach must be provided as well as validation input and output data. If a pre-trained model has been used, it must be clearly identifiable. For both supervised and unsupervised machine learning applications, the provided example or validation data must not be part of the training and testing data.

**Recommended.** To facilitate the reproduction and validation of results from either models trained from scratch or pre-trained models that were re-trained, the full training and testing data and any training metadata (e.g., training time) should be made available. The code used for training the model should be provided. Code, as well as data, should be provided via public repositories (European Organization For Nuclear Research and OpenAIRE 2013). The authors should discuss and ideally test how well the model has performed and show any limitations of the used machine learning approach on their data. The application of machine learning models will particularly benefit from being deployed in a cloud-hosted format or via software containers.

**Ideal.** Further standardization promotes ease of reproduction and validation by the scientific community by making use of emerging online platforms. Thus, models could be created conforming to standardized formats (e.g., [Model zoo](#)) if they become more readily available in the future.

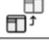

*Figure 8. Checklist for publication of image analysis workflows.*

## Discussion

Herein we have presented recommendations in the form of checklists to increase the understandability and reproducibility of published image figures and image analyses. While our checklists were initially intended for bioimages from light microscopes, we do believe that its many principles are applicable more widely. Our checklists include recommendations for image formatting, annotation, color display, and data availability , which at the minimal level can largely be achieved with commercial or open-source software (e.g., 'include scale bar'). Likewise, the minimal suggestions for image analysis pipelines can be implemented readily with today's options (e.g., code repositories). We believe that, once included in microscopy core facility training and microscopy courses, and introduced as guidelines from publishers, the recommendations will present no additional burden. On the contrary, transparent requirements for publishing images and progress monitoring checklists will ease the path from starting a microscopy experiments to producing reproducible (Baker, 2016) understandable image figures for all scientists.

Recommendations extending the "Minimal" level are introduced in the "Recommended" and "Ideal" reporting levels and at times go beyond what is easy to implement with standard tools today. They are meant to encourage a continuous strive towards higher quality standards in image publishing. Before all of these advanced standards can become a new norm, technologies, software and research infrastructure must still be improved. At present no image database is used widely enough to become a go-to solution, although dedicated resources exist and are slowly getting traction, and publishers are experimenting with parallel solutions (e.g., EMBO source data). Also, while funding agencies increasingly require data to be deposited in repositories, few guidelines are provided for publishing terabytes to petabytes of raw data. While publishers may mandate data deposition or availability, they are not always reviewing its implementation. Combined with a lack of recognition of efforts put into publishing original image data, scientists are often discouraged to make data openly available. Commercial solutions for data storage are increasingly becoming available. For instance the AWS Open Data has already been used to host image data (https://registry.opendata.aws/cellpainting-gallery/) and we believe that, ultimately, images presented in most publications should be linked to a losslessly compressed image amenable to re-analysis.

The checklists and recommendations for image analysis will naturally be dynamic and require regular updates to reflect new developments in this active research domain. Moreover, it is possible that generation of publication quality images will also become a standardized 'workflow' in and of itself. It was previously suggested that images should be processed through scripting, with every step, from microscope output to published figure, stored in a metadata file (Miura and Norrelykke, 2021). Another challenge is the continuous availability of  image analysis software and workflows, which requires software maintenance and updates to stay usable.

We envision that the present checklists will be continuously updated by the scientific community and adapted to future requirements and unforeseen challenges. Future work of the Image Analysis and Visualization Working Group will be to, in alliance with similar initiatives such as NEUBIAS (Cimini et al., 2020; Martins et al., 2021) and BINA, develop educational materials and tutorials based on the presented checklists and to continuously lobby to integrate its contents in general resources for better images (Collins et al., 2017). We ask that all readers consider how their work will be seen and used in the future and join us in building a stronger scientific foundation for everyone. The presented checklists, version 1.0, will already make images in publications more accessible, understandable and reproducible, providing a valuable resource that may be used to build a solid foundation within today's research that will benefit future science and scientists.

## Author Affiliations

[1]Fondazione Human Technopole, Viale Rita Levi-Montalcini 1, 20157 Milano, Italy
[2]Present address: Leibniz-Forschungsinstitut für Molekulare Pharmakologie (FMP), Robert-Rössle-Str. 10, 13125 Berlin, Germany
[3]Department of Biomedical Engineering, University of Wisconsin-Madison, Madison, WI, 53706, USA.
[4]Max Planck Institute of Immunobiology and Epigenetics, 79108 Freiburg, Germany
[5]Medical BioSciences department, Radboud University Medical Centre, Nijmegen, Netherlands
[6]Laboratory for Molecular mechanics of cell adhesions, Pontificia Universidad Católica de Chile Santiago
[7]Osaka University, Graduate School of Engineering Science, Japan
[8]Euro-BioImaging ERIC, Bio-Hub, Meyerhofstr. 1, 69117 Heidelberg, Germany
[9]Carl Zeiss AG, Carl-Zeiss-Straße 22, 73447 Oberkochen, Germany
[10]BioVoxxel, Scientific Image Processing and Analysis, Eugen-Roth-Strasse 8, 67071 Ludwigshafen, Germany
[11]Nanophotonics and BioImaging Facility at INL, International Iberian Nanotechnology Laboratory, 4715-330, Portugal
[12]Phi Optics, Inc., 1800 S. Oak St, Ste 106, Champaign, IL 61820, USA
[13]Biochemistry Center Heidelberg, Heidelberg University, Germany
[14]Imaging Platform, Broad Institute, Cambridge, MA 02142
[15]i3S, Instituto de Investigação e Inovação Em Saúde and INEB, Instituto de Engenharia Biomédica, Universidade do Porto, Porto, Portugal
[16]Translational Research Institute, Queensland University of Technology, 37 Kent Street, Woolloongabba, QLD 4102, Australia
[17]School of Biomedical Sciences, Faculty of Health, Queensland University of Technology, Brisbane, QLD 4059, Australia
[18]TissueGnostics GmbH, 1020 Vienna, Austria
[19]Department of Medical Physics and Biomedical Engineering, University of Wisconsin-Madison, Madison, WI, 53706, USA
[20]Centre for Cellular Imaging Core Facility, Sahlgrenska Academy, University of Gothenburg, Sweden
[21]Allen Institute for Cell Science, Seattle, WA, USA
[22]Friedrich Miescher Institute for Biomedical Research, Basel, Switzerland
[23]RNA Therapeutics Institute, University of Massachusetts Chan Medical School, Worcester, MA 01605, USA
[24]University of Oklahoma, Norman, OK, USA
[25]Advanced Light Microscopy Unit, Centre for Genomic Regulation, Barcelona, Spain
[26]European Molecular Biology Laboratory, European Bioinformatics Institute (EMBL-EBI), Hinxton, UK
[27]Centre for Cell Imaging, The University of Liverpool, UK
[28]Department of AI research, Kapoor Labs, Paris, 75005, France
[29]MD Anderson Cancer Center, Houston, TX, USA
[30]MIA Cellavie Inc., Montreal, QC Canada
[31]Division of Drug Discovery and Safety, Cell Observatory, Leiden Academic Centre for Drug Research, Leiden University, 2333 CC Leiden, The Netherlands
[32]Karolinska Institutet, Hälsovägen 7C, 14157, Huddinge, Sweden
[33]Key Laboratory for Modern Measurement Technology and Instruments of Zhejiang Province, College of Optical and Electronic Technology, China Jiliang University, Hangzhou, China
[34]Advanced Imaging Facility, Instituto Gulbenkian de Ciência, Oeiras 2780-156 – Portugal
[35]Bioimage Analysis & Research, 69127 Heidelberg/Germany
[36]MicRoN Core, Harvard Medical School, Boston, MA, USA

[37]Life Imaging Center, Signalling Research Centres CIBSS and BIOSS, University of Freiburg, Germany
[38]Bio-Imaging Resource Center, The Rockefeller University, New York, NY USA
[39]Baker Institute Microscopy Platform, Baker Heart and Diabetes Institute, Melbourne, VIC, 3004, Australia
[40]School of Physics, Engineering and Technology, University of York, Heslington, YO10 5DD, UK
[41]King Abdullah International Medical Research Center (KAIMRC), Medical Research Core Facility and Platforms (MRCFP), King Saud bin Abdulaziz University for Health Sciences (KSAU-HS), Ministry of National Guard Health Affairs (MNGHA), Riyadh 11481, Saudi Arabia
[42]Light Microscopy Facility, Max Planck Institute of Molecular Cell Biology and Genetics Dresden, Pfotenhauerstrasse 108, 01307 Dresden, Germany
[43]ariadne.ai (Germany) GmbH, 69115 Heidelberg, Germany
[44]Space Biosciences Division, NASA Ames Research Center, Moffett Field, CA, 94035, USA
[45]BioImaging & Optics Platform (BIOP), Ecole Polytechnique Fédérale de Lausanne (EPFL), Faculty of Life sciences (SV), CH-1015 Lausanne
[46]Microscopy & BioImaging Consulting, Image Processing & Large Data Handling, Tobias-Hoppe-Strassse 3, 07548 Gera, Germany
[47]Max Planck Institute of Biochemistry, Am Klopferspitz 18, 82152 Martinsried, Germany
[48]Imaging and Morphology Support Core, Oregon National Primate Research Center - (ONPRC - OHSU West Campus), Beaverton, Oregon 97006, USA.
[49]Program in Molecular Medicine, University of Massachusetts Chan Medical School, Worcester, MA, 01605, USA
[50]Department of Pathology and Laboratory Medicine, Microscopy Imaging Center (RRID# SCR_018821), Center for Biomedical Shared Resources, University of Vermont, Burlington, VT 05405 USA
[51]Centre for Bioimage Analysis, EMBL Heidelberg, Meyerhofstr. 1, 69117 Heidelberg, Germany
[52]NCT-UCC, Medizinische Fakultät TU Dresden, Fetscherstrasse 105, 01307 Dresden/Germany

## Author CREDIT statements

C. S. Conceptualization, Methodology, Writing Original Draft, Visualization, Supervision, Project administration.
M. S. N. Conceptualization, Methodology, Writing Original Draft, Visualization, Writing - Review and Editing.
S. A. Endorsement & Reviewing.
G.-J. B. Endorsement & Reviewing.
C. B. Critical review - commentary in pre-writing stage - Review & Editing.
J. Bischof Endorsement & Reviewing
U. B. Conceptualization, Methodology, Writing - Review & Editing, Endorsement.
J. Brocher. Endorsement, Writing - Review & Editing, Visualization.
M. T. C. Conceptualization, Methodology, Endorsement, Writing - Review & Editing.
C. C. Endorsement & reviewing.
J. C. Writing - Review & Editing.
B. A. C. Writing - Review & Editing, Endorsement.
E. C.-S. Endorsement & reviewing.
M. E. Writing Review and Editing.
R. E. Endorsement & reviewing.
K. E. Endorsement & reviewing.
J. F.- R. Endorsement & reviewing.
N. G. Endorsement & reviewing.

L. G. Endorsement & reviewing
D. G. Writing - Review and Editing.
T. G. Writing - Review and Editing.
N. H. Endorsement, review & editing.
M. Hammer Review and Editing.
M. Hartley Endorsement & reviewing.
M. Held Endorsement & reviewing.
F. J. Endorsement & reviewing.
V. K. Writing - Review and Editing.
A. A. K. Endorsement & reviewing and Editing
J. L. Endorsement & reviewing.
S. LeD. Endorsement & reviewing.
S. LeG. Writing - Review & Editing.
P. L. Endorsement & reviewing.
G. G. M. Conceptualization, Methodology, Writing - Review & Editing.
A. M. Reviewing and Editing, Endorsement.
K. M. Conceptualization, Methodology, Writing - Review and Editing.
P. M. L. Endorsement & Reviewing.
R. N. Conceptualization, Supervision, Project administration, Endorsement.
A. N. Endorsement & Reviewing.
A. C. P. Conceptualization, Methodology, Writing - Review and Editing.
A. P.-D. Writing - Review and Editing, Visualization.
L. P. Writing - Review and Editing.
R. A. Endorsement & Reviewing.
B. S.-D. Endorsement & Reviewing.
L. S. Endorsement & Reviewing.
R. T. S. Endorsement & reviewing.
A. S. Review and Editing.
O. S. Endorsement, Writing - Review & Editing.
V. P. S. Endorsement, reviewing.
M. S. Endorsement & reviewing, Writing – Review and Editing.
S. S. Resources, Writing - Review and Editing.
C. S.-D.-C. Conceptualization, Endorsement, Validation.
D. J. T. Writing - Review and Editing.
C. T. Conceptualization, Methodology, Writing - Review & Editing.
H. K. J. Conceptualization, Methodology, Writing Original Draft, Visualization, Project administration, Supervision

## Data Availability

The checklists can be downloaded as printable files here: https://doi.org/10.5281/zenodo.7642559

The companion Jupyter Book can be found here: https://quarep-limi.github.io/WG12_checklists_for_image_publishing/intro.html

## Acknowledgements

We acknowledge the support and discussions with all our colleagues and QUAREP-LiMi Working Group 12 members. We additionally thank Oliver Biehlmeier, Martin Bornhäuser, Oliver Burri, Caron Jacobs, Alex Laude, Kenneth Ho, Rocco D'Antuono for further feedback and discussions on this manuscript and their endorsement of the checklists. Icons: designed by the authors and from Fontawesome.com; Images: ImageJ sample images (Schneider et al., 2012), (Jambor et al., 2015) and (Sarov et al., 2016).

C.S. was supported by a grant from the Chan Zuckerberg Initiative napari Plugin Foundation #2021-24038.
C.B. was supported by grants ANID (PIA ACT192015; Fondecyt 1210872; Fondequip EMQ210101; Fondequip EQM210020) and PUENTE-2022-13 from Pontificia Universidad Católica de Chile.
B.A.C. was funded by NIH P41 GM135019 and grant 2020-225720 from the Chan Zuckerberg Initiative DAF, an advised fund of the Silicon Valley Community Foundation.
F.J. was supported by AI4Life (European Unions's Horizon Europe research and innovation programme, #101057970) and by the Chan Zuckerberg Initiative napari Plugin Foundation #2021-240383 and #2021-239867.
M.Hammer and D.G. were supported by NSF award 1917206 and NIH award U01 CA200059.
R.N. was supported by grant NI 451/10-1 from the German Research Foundation and grant 03TN0047B "FluMiKal" from the German Federal Ministry for Economic Affairs and Climate Action.
A.P-D. was supported by EPSRC grant EP/W024063/1
C.S.D.C was supported by grant #2019-198155 (5022) awarded by the Chan Zuckerberg Initiative DAF, an advised fund of Silicon Valley Community Foundation, as part of their Imaging Scientist Program. She was also funded by NIH grant #U01CA200059.
C.T. was supported by a grant from the Chan Zuckerberg Initiative DAF, an advised fund of Silicon Valley Community Foundation (grant number 2020- 225265).
H.K.J. was supported by MSNZ funding of the Deutsche Krebshilfe.

# Glossary

| Term | Description |
| --- | --- |
| Image | Used here for image data from a microscope experiment, principles described may also apply to medical images, electron microscopy images. |
| Original image | Output files/source image data of the microscope; depending on microscope type and the vendor these may be essentially "raw", i.e., what is visible through the ocular, or pre-processed. |
| Workflow | A series of image processing and analysis steps to generate a meaningful result for only a specific application, without reusability in mind. The individual steps typically use existing image analysis components. A workflow exists usually as a script or plugin within a software platform or as a stand-alone. See also workflow template |
| Image analysis component | Computer vision methods and algorithms that are available as functions or classes in software platforms for image analysis |
| Software platform/library | Software that bundles many algorithms, tools, and workflows (e.g., Fiji, CellProfiler). |
| Workflow template | A workflow that is engineered such that it can be reused for more applications and different users. Typically is created with more flexibility and accessibility in mind. Thus, provides more options to modify for a different use case and exposes settings in an easy-to-use manner (e.g., GUI). |
| GUI | Graphical user interface |
| Machine learning model | A program that makes a decision (classifier) or returns an output (regression) based on some input, with the ability to process previously unseen data. |
| Channel adjustment | Change to the brightness, contrast, or gamma correction. |
| Contrast | The difference between the brightest and darkest pixels in an image. |
| Supervised machine learning | Training a machine learning model with labeled data, for example the inputs for training have been previously classified by a human. |
| Unsupervised machine learning | Training a machine learning model with unlabeled data, often to perform tasks such as clustering |
| Deep learning | Machine learning using deep neural networks. |
| Ground truth | Labeled data. While often described as ground truth, mistakes are often made, especially in large data sets, and should not be assumed to be the actual truth. |
| Software containers | A versioned, reproducible, and reusable computing system (such as an operating system visualizer such as Docker (https://www.docker.com/) or Singularity (https://docs.sylabs.io/guides/3.5/user-guide/introduction.html) or an otherwise reusable virtual machine system) that allows arbitrary numbers of users to access one or more software tools in a controlled and defined environment. |

# Supplemental materials

*Supplemental figure 1.* Alternative layout Checklist for image Publishing

## Checklist for image publishing

### Image format

| | | |
|---|---|---|
| Focus on relevant image content (e.g. crop, rotate, resize) | ☐ | Minimal |
| Separate individual images | ☐ | |
| Show example image used for quantifications | ☐ | |
| Indicate position of zoom-view/inset in full-view/ original image | ☐ | |
| Show images of the range of described phenotype | ☐ | |

### Image colors and channels

| | | |
|---|---|---|
| Annotation of channels (staining, marker etc.) visible | ☐ | Minimal |
| Adjust brightness/contrast, report adjustments, use uniform color-scales | ☐ | |
| Image comparison: use same adjustments | ☐ | |
| Multi-color images: accessible to color blind | ☐ | |
| Channel color high visibility on background | ☐ | |
| Provide grey-scale for each color channel | ☐ | Recommended |
| Provide color scales for intensity values (greyscale, color, pseudo color…) | ☐ | |
| Pseudo-colored images: additionally provide greyscale version for comparison. | ☐ | Ideal |
| Gamma adjustments: additionally provide linear-adjusted image for comparison | ☐ | |

### Image annotation

| | | |
|---|---|---|
| Add scale information (scale bar, image length) | ☐ | Minimal |
| Explain all annotations | ☐ | |
| Legible annotations (point size, color) | ☐ | |
| Annotations should not obscure key data | ☐ | |
| Annotate image dimensions (z-distance in image stacks), image pixel size, imaging intervals (time-lapse in movies) | ☐ | Recommended |
| Annotate imaging details e.g., exposure time | ☐ | Ideal |

### Image availability

| | | |
|---|---|---|
| Images are shared (lossless compression/microscope images) | ☐ | Minimal |
| Image files are freely downloadable (public database) | ☐ | Recommended |
| Image files are in dedicated image database (added value database or image archive) | ☐ | Ideal |

*Supplemental figure 2.* Alternative layout Checklist for image Analysis Publishing

## Checklists for publication of image analysis workflows

### Established workflows

| Icon | Item | Level |
|---|---|---|
| | Cite workflow & platform | Minimal |
| | Key settings | |
| | Example data | |
| | Manual ROIs | |
| 1.9.3 | Exact version | Recommended |
| | All settings | |
| | Public example | |
| | Document usage (e.g. screen recording or tutorial) | Ideal |
| | Cloud hosted or container | |

### Novel workflows

| Icon | Item | Level |
|---|---|---|
| | Cite components & platform | Minimal |
| | Describe sequence | |
| | Key settings | |
| | Example data & code | |
| | Manual ROIs | |
| 1.9.3 | Exact versions | |
| | All settings | Recommended |
| | Public example data & code | |
| | Rationale | |
| | Limitations | |
| | Screen recording or tutorial | Ideal |
| | Easy install & usage, container | |

### Machine learning workflows

| Icon | Item | Level |
|---|---|---|
| | Cite original method | Minimal |
| | Access to model | |
| | Example or validation data | |
| | Train, test & metadata | Recommended |
| | Code available | |
| | Limitations | |
| | Cloud hosted or container | |
| | Standardized format | Ideal |

***Supplemental figure 3.*** *Overview of current repositories that accept image data. (Cimini, 2023)*

| Repository | Zenodo | Figshare / Figshare+ | Dryad | Bioimage Archive | Image Data Repository | Broad Bioimage Benchmark Collection | The Cell Painting Gallery |
|---|---|---|---|---|---|---|---|
| URL | https://zenodo.org/ | https://figshare.com/ | https://datadryad.org/ | https://www.ebi.ac.uk/bioimage-archive/ | https://idr.openmicroscopy.org/ | https://broad.io/BBBC | https://broad.io/cellpaintinggallery |
| Qualifications | "All the digital artefacts" | Research data outputs | Non-human identifiable data of any kind that authors are willing to make CC0 | Non-medical, non-Electron Microscopy images of any kind | "Reference image datasets" - complete, can be associated with other resources, likely to be re-analyzed | Image sets with descriptions and ground truth | Microscopic image sets suitable for image-based profiling |
| Also non-image data? | Yes | Yes | Yes | Not directly (BioStudies) | Not directly | No | Sometimes |
| Size limit? | 50 GB per collection (soft) | 20 GB Figshare, 5TB Figshare+ (soft) | 300 GB (soft) | No | 1 TB (soft) | No | No |
| Cost to depositor? | No ("donations encouraged") | Free to $20 GB, $395 to 100 GB, $585/250 GB beyond | $120, + $50/every 10 GB over 50 GB (some funders provide sponsorship) | No | No | No | No |
| Strictness of metadata requirements | None | Low | Low | Medium | High | Medium | High |